\documentclass[preprint,showpacs,preprintnumbers,amsmath,amssymb]{revtex4}

\usepackage{graphicx}
\usepackage{dcolumn}
\usepackage{bm}

\begin{document}

\title{\large Proximity induced superconductivity by Bi in topological
$Bi_2Te_2Se$ and $Bi_2Se_3$  films:
Evidence for a robust zero energy bound state possibly due to Majorana fermions }

\author{G. Koren}
\email{gkoren@physics.technion.ac.il} \affiliation{Physics
Department, Technion - Israel Institute of Technology Haifa,
32000, ISRAEL} \homepage{http://physics.technion.ac.il/~gkoren}

\author{T. Kirzhner}
\affiliation{Physics Department, Technion - Israel Institute of
Technology Haifa, 32000, ISRAEL}

\author{E. Lahoud}
\affiliation{Physics Department, Technion - Israel Institute of
Technology Haifa, 32000, ISRAEL}

\author{K. B. Chashka}
\affiliation{Physics Department, Technion - Israel Institute of
Technology Haifa, 32000, ISRAEL}

\author{A. Kanigel}
\affiliation{Physics Department, Technion - Israel Institute of
Technology Haifa, 32000, ISRAEL}

\date{\today}
\def\bfig {\begin{figure}[tbhp] \centering}
\def\efig {\end{figure}}

\normalsize \baselineskip=8mm  \vspace{15mm}


\begin{abstract}

Point contact conductance measurements on topological $Bi_2Te_2Se$ and $Bi_2Se_3$ films reveal a signature of superconductivity below 2-3 K. In particular, critical current dips and a robust zero bias conductance peak are observed. The latter suggests the presence of zero energy bound states which could be assigned to Majorana fermions in an unconventional topological superconductor. We attribute these novel observations to proximity induced local superconductivity in the films by small amounts of superconducting Bi inclusions or segregation to the surface, and provide supportive evidence for these effects.

\end{abstract}

\maketitle

\section{Introduction }
\normalsize \baselineskip=6mm  \vspace{6mm}

In recent years a set of insulators with strong spin-orbit coupling and time reversal symmetry has been identified as comprising topological insulating
electronic phases \cite{KaneRMP}. Basically, ideal topological insulators (TOI) have a finite energy gap and are insulating in the bulk while on the surface (or edges in 2D) they are gapless and have protected conducting states. These states had been observed in stoichiometric bulk materials such as $Bi_2Se_3$ and $Bi_2Te_3$ by angular resolved photoemission spectroscopy (ARPES) measurements \cite{ARPES}. However, in topological thin films of these chalcogenides, generally, many Se vacancies are present, which make these materials more conducting and even metallic. Further doping of these materials as was recently done in $Bi_2Se_3$ by copper intercalation, renders them superconducting with a transition temperature $T_c$ in the range of 3-4 K \cite{Hor,Ando}. This could result in a phase of matter which is a topological superconductor (TOS) whose hallmark signature would be the presence of Majorana fermions (MF). These MF  will lead to the appearance of a clear zero bias conductance peak (ZBCP) in conductance spectra of the TOS, which reflects their zero energy bound state nature \cite{Ando}. Sasaki \textit{et al}. had actually observed such a ZBCP in point contact measurements on superconducting $Cu_xBi_2Se_3$ single crystals, and concluded from comprehensive theoretical considerations that they are due to MF which supports the TOS scenario \cite{Ando}. Similar kind of measurements were performed in parallel by our group, and the results support this conclusion \cite{Tal}. Another way to obtain superconductivity in a TOI is by means of the proximity effect with a known superconductor. Yang \textit{et al.} have used this method and observed a very narrow ZBCP ($\sim$0.1 mV) in junctions of Sn ($T_c\sim 3.8$ K) and $Bi_2Se_3$ single crystal flake at low temperatures \cite{LiLu}. In the present study we used proximity induced superconductivity in topological insulator films, and looked for the Majorana fermion signature in their point contact conductance spectra. Unexpectedly, we found that TOS and MF were observed not only in bilayers of superconducting Bi and $Bi_2Te_2Se$, but also in pure $Bi_2Te_2Se$ and $Bi_2Se_3$ films. We concluded that due to the large amount of Se vacancies in these films, Bi segregation to the surface as well as Bi inclusions led to the observed results.\\

\section{Preparation and characterization of the films }
\normalsize \baselineskip=6mm  \vspace{6mm}

\subsection{ $Bi_2Se_3$ films and $Bi-Bi_2Te_2Se$ bilayers}

\begin{figure} \hspace{-20mm}
\includegraphics[height=8cm,width=13cm]{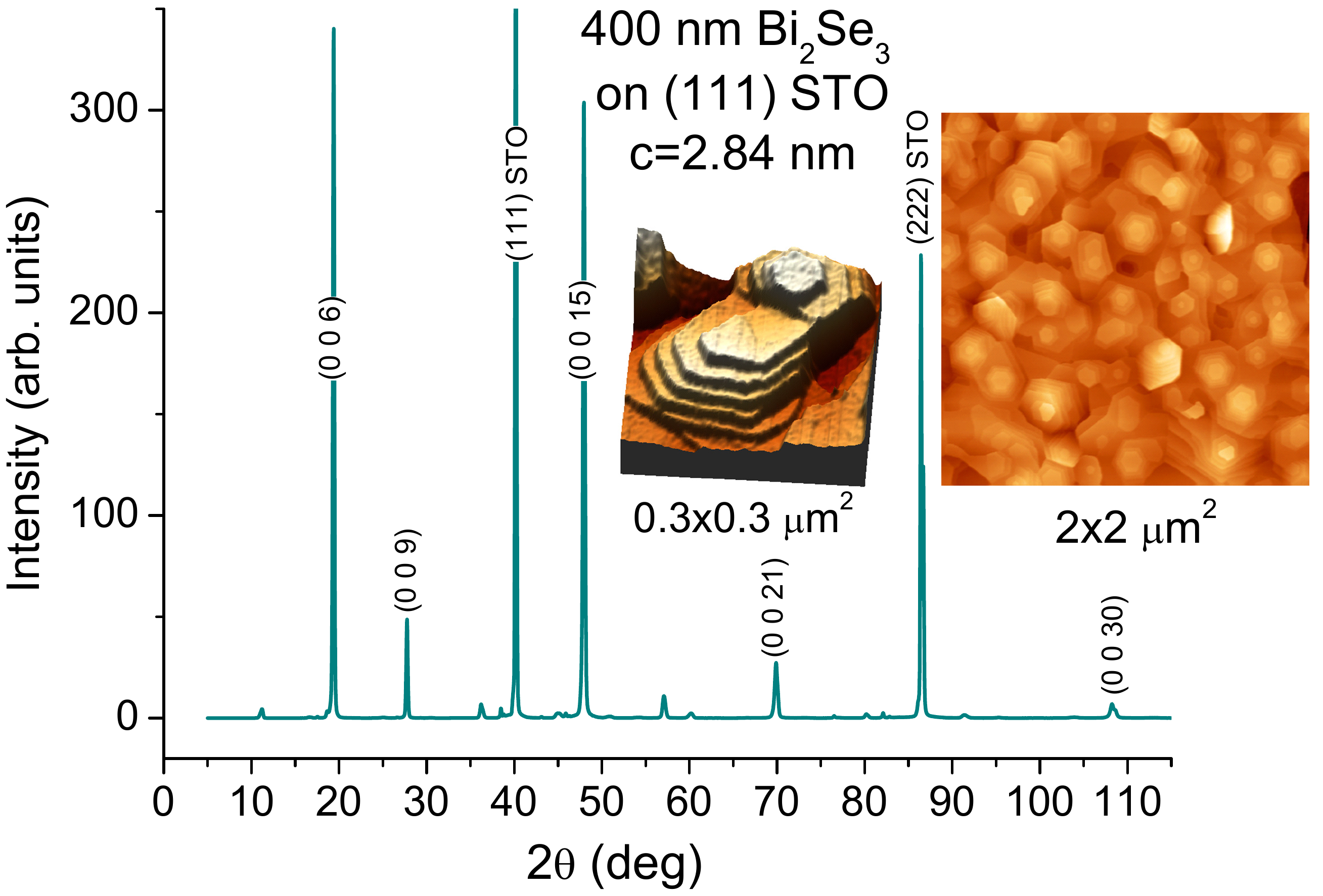}
\vspace{-0mm} \caption{\label{fig:epsart} (Color online) X-ray diffraction results of a 400 nm thick $Bi_2Se_3$ film deposited on (111) $SrTiO_3$ wafer. The insets show AFM images of the surface morphology of this film. }
\end{figure}

First we optimized the deposition temperature of the $Bi_2Se_3$ films by checking their structure using x-ray diffraction, and surface morphology by atomic force microscopy (AFM). Deposition was done by laser ablation under vacuum from a stoichiometric polycrystalline $Bi_2Se_3$ target. We found that in the temperature range of 300-600 $^0$C, the best temperature for the crystallization of the film was 400 $^0$C. We note that this is the heating block temperature to which the wafer was clamped, and this led to an actual deposition temperature of about 200 $^0$C on the surface of the film. Then we optimized the laser fluence on the target, in order to obtain the correct $Bi_2Se_3$ phase of the films. At high laser fluence of about 1 $J/cm^2$, Bi rich films of $Bi_4Se_3$ were obtained and only when the laser fluence was reduced to 0.4 $J/cm^2$ the desired $Bi_2Se_3$ phase was achieved. The lower fluence on the target was needed in order to have less energetic Se atoms and ions in the laser ablated plume, which enhanced their sticking to the film and prevented higher Se losses. The resulting optimized film structure is given by the x-ray diffraction patterns of Figs. 1 and 2 on two different substrates, (111) $SrTiO_3$ and (100) $MgO$, together with AFM images of the surface morphology of the first film. As can be seen in these figures, on both wafers the growth was with the c-axis normal to the wafer with c=2.84 nm. The AFM images in Fig. 1 show that the film on the (111) $SrTiO_3$ wafer was also epitaxial, with clear and ordered in-plane hexagonal structure. Fig. 3 shows x-ray diffraction results measured on a $Bi-Bi_2Te_2Se$ bilayer on (100) $LaAlO_3$ which was used in the proximity effect studies. The results are basically similar to those of Figs. 1 and 2 of the $Bi_2Se_3$ films, but with the additional (0,0,3n) peaks of the hexagonal Bi phase (R$\_$3m:H symmetry with c=1.18 nm). Some other peaks which are unaccounted for in this figure can be due to the other phases of Bi.\\

\begin{figure} \hspace{-20mm}
\includegraphics[height=7cm,width=12cm]{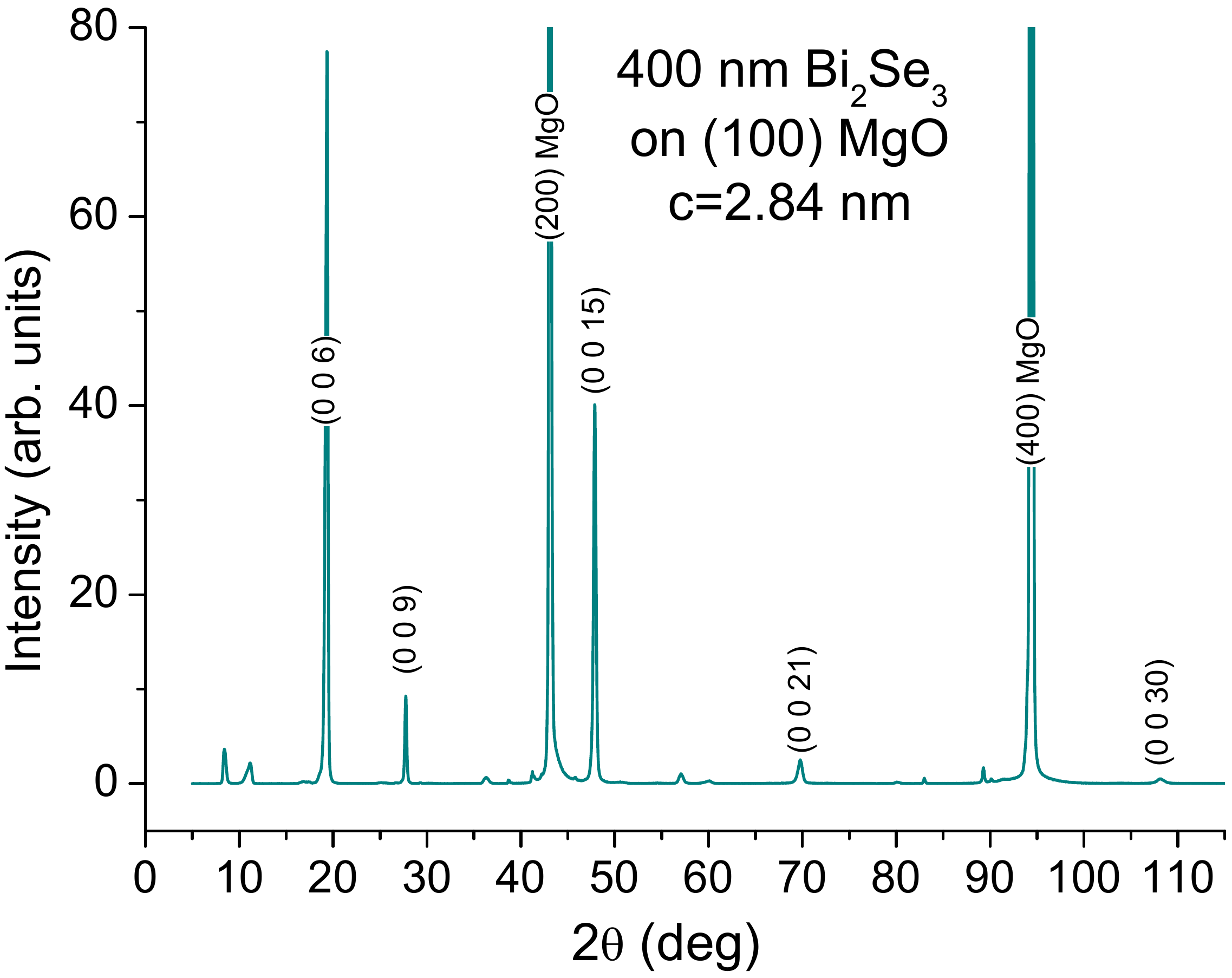}
\vspace{-0mm} \caption{\label{fig:epsart} (Color online) X-ray diffraction results of a 400 nm thick $Bi_2Se_3$ film deposited on (100) $MgO$ wafer. }
\end{figure}

\begin{figure} \hspace{-20mm}
\includegraphics[height=7cm,width=12cm]{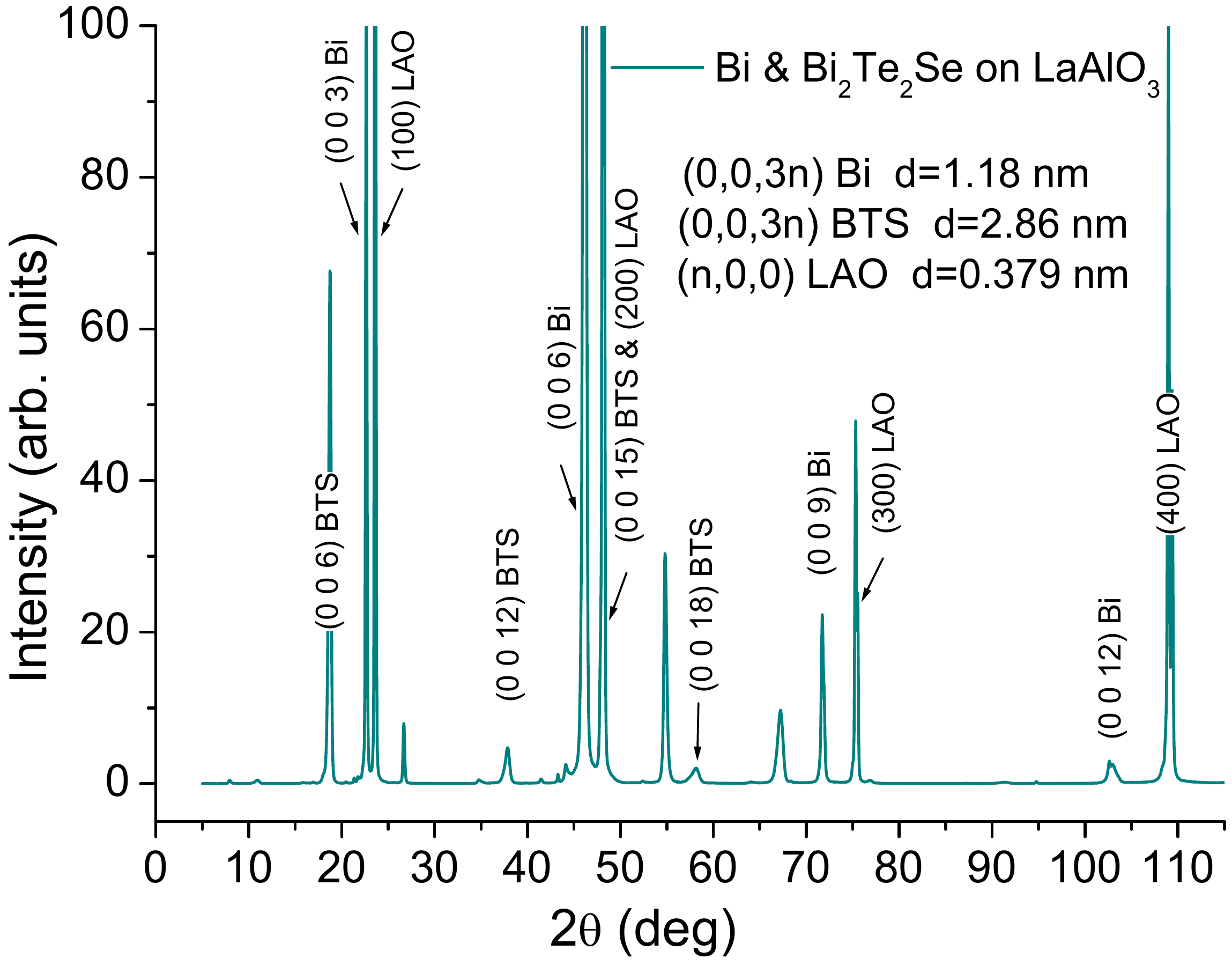}
\vspace{-0mm} \caption{\label{fig:epsart} (Color online) X-ray diffraction results of a 100 nm thick Bi layer on top of a 700 nm thick $Bi_2Te_2Se$ (BTS) film deposited on (100) $LaAlO_3$ wafer.}
\end{figure}

\begin{figure} \hspace{-20mm}
\includegraphics[height=9cm,width=13cm]{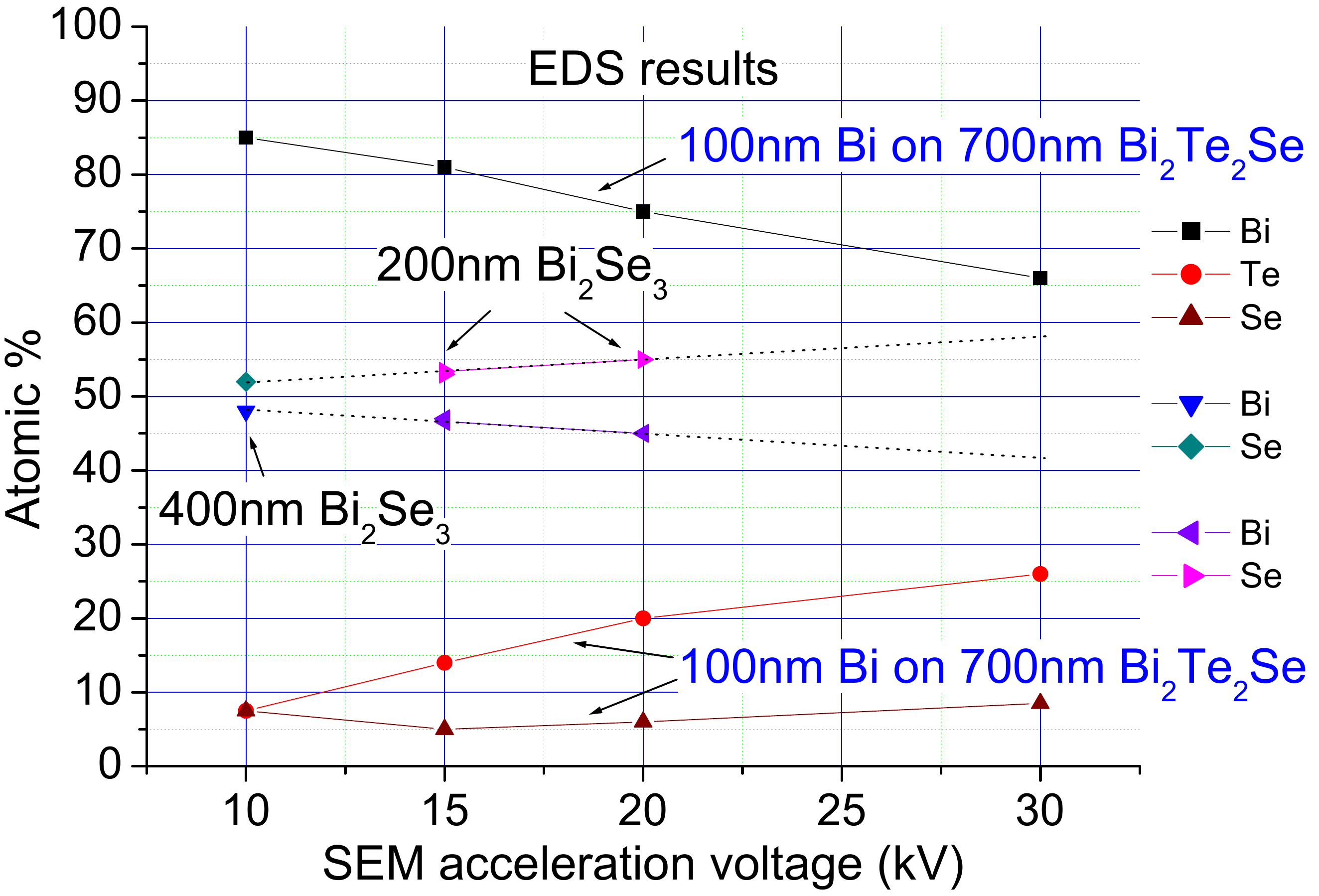}
\vspace{-0mm} \caption{\label{fig:epsart} (Color online) Energy dispersive spectroscopy (EDS) results in atomic \% versus the accelerating voltage of the scanning electron microscope (SEM) of two $Bi_2Se_3$ films on (111) $SrTiO_3$ and (100) $MgO$,  and of the $Bi-Bi_2Te_2Se$ bilayer on (100) $LaAlO_3$. }
\end{figure}

\subsection{Se loss and Bi segregation to the surface of the films}

Fig. 4 shows energy dispersive spectroscopy (EDS) data on two $Bi_2Se_3$ films and on the $Bi-Bi_2Te_2Se$ bilayer. The data is given by calibrated atomic percents as a function of the scanning electron microscope (SEM) accelerating voltage. This mimics qualitatively the atomic concentration depth profile of the films, but is clearly less accurate than concentration depth profiles that could be obtain by Auger spectroscopy or secondary ion mass spectrometry (SIMS). Nevertheless, the EDS data shows that in the bare $Bi_2Se_3$ films, the Se:Bi ratio is lower than the expected 3/2, but is approaching systematically this ratio with increasing acceleration voltage in the SEM (see the two dotted extrapolation lines in Fig. 4). Qualitatively, this indicates that the surface layer is Bi rich, and that this can originate in segregation of Bi to the surface due to losses of the volatile Se from the film. In the $Bi-Bi_2Te_2Se$ bilayer the picture is even more complex. Since the top layer is Bi, at low SEM voltages one sees mostly Bi. With increasing voltage though, the electrons penetrate deeper into the bilayer and the emitted x-rays emanate also from deeper layers. Fig. 4 shows that for the bilayer the Bi concentration goes down with accelerating voltage while the Te and Se concentrations go up. More so for the Te than for the Se, since the Se is much more volatile than Te. Even at 30 kV, the Te:Se ratio is still lower than 2. Note that the high Bi concentration here is due mostly to the Bi cap layer, and thus can not be compared with the $Bi_2Te_2Se$ formula ratios.\\

\begin{figure} \hspace{-20mm}
\includegraphics[height=9cm,width=13cm]{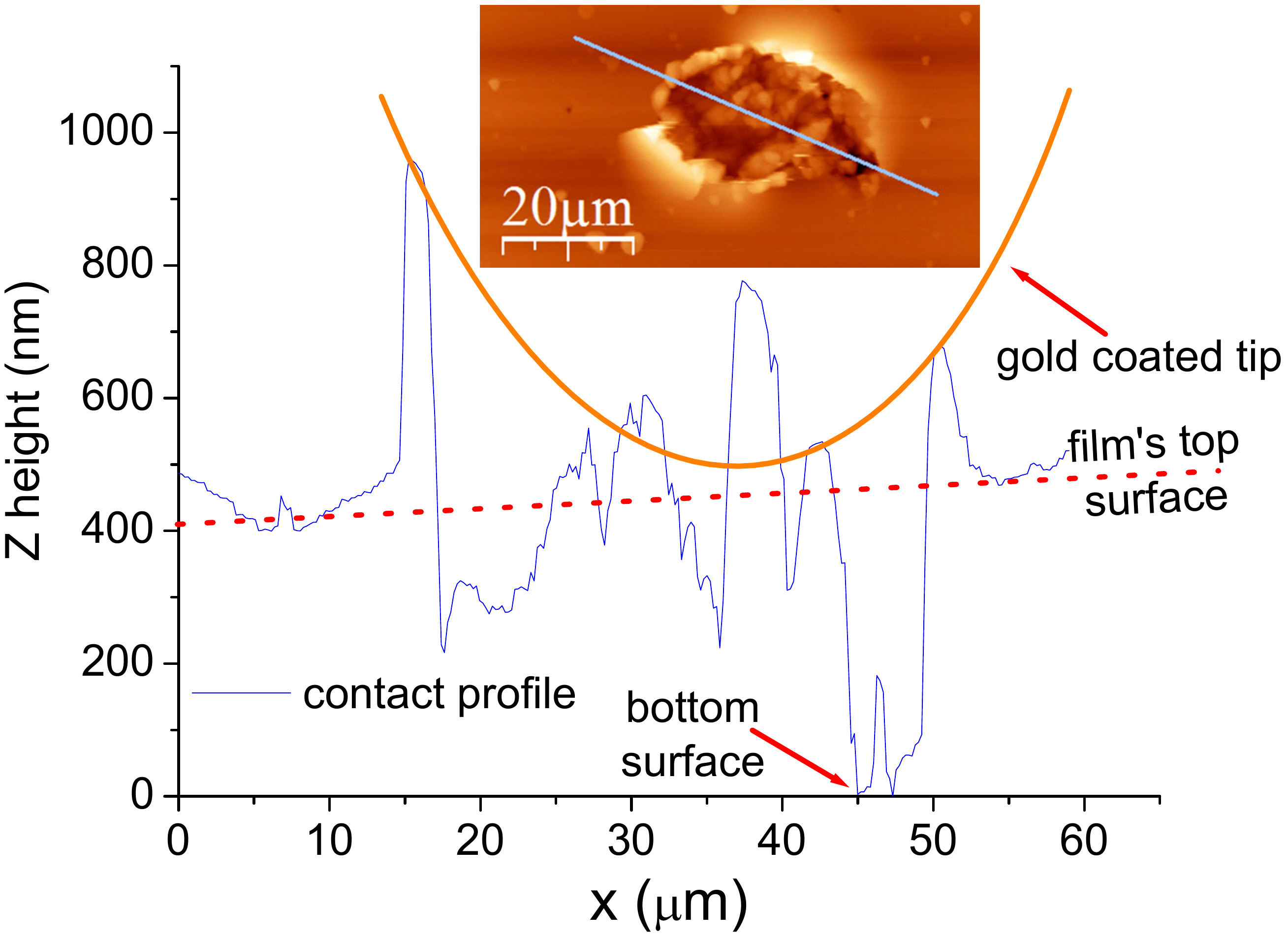}
\vspace{-0mm} \caption{\label{fig:epsart} (Color online) A depth profile of a contact hole in a 400 nm thick $Bi_2Se_3$ film on (111) $SrTiO_3$ along the line shown in the AFM image of the inset (same film as in Fig. 1). A schematic drawing of the gold coated tip is also outlined.}
\end{figure}

\subsection{The point contact junctions}

An array of forty gold coated spring loaded spherical tips of 0.5 mm diameter was pressed onto the films and produced 40 point contact junctions on each wafer. We used 4-probes to measure the resistance versus temperature of the films, and 2-probes to measure their conductance spectra. In all, the conductance spectra were measured on 60 combinations of 2 contacts on each wafer which constituted a reasonable statistics of the 40 point contacts response. Repeated measurement on fresh areas of the same film provided more data with the same statistics. The 60 combinations of 2 contacts also allowed for separation of the small number of "active" contacts which were affected by superconductivity ($<10\%$) from the normal resistive ones. The voltage scale of the conductance spectra of each contact in this study V=$V(contact)$ was always calibrated by using the equation $V(measured)=2V(contact)+V(film)+V(leads)$ which takes into account the voltage drops on the two contacts in series, on the film in between the contacts, and on the leads to the sample. Fig. 5 shows a depth profile of a typical contact hole left behind after the contact tip was removed in the 400 nm thick $Bi_2Se_3$ film on (111) $SrTiO_3$ of Fig. 1.  Since the film is brittle, the pressure applied to the contact determined its penetration depth into the film. The profile was taken along the line shown in the AFM image of the inset. The top surface of the film is outlined by the dotted line in this figure while in some points the hole reaches the substrate at its bottom surface. A schematic drawing of the gold coated tip is also shown, in a position of just losing contact with the film when it was pulled out. Clearly, the spring loaded gold coated tip pierced the film, pushed some material to its sides and also pulled some flakes upward, above the surface of the film, when it was pulled out. We thus conclude that the contact between the tip and the film occurs across many much smaller areas (or points) than the overall hole area.\\

\subsection{Conductance spectra of Au-Bi point contact junctions }

\begin{figure} \hspace{-20mm}
\includegraphics[height=9cm,width=13cm]{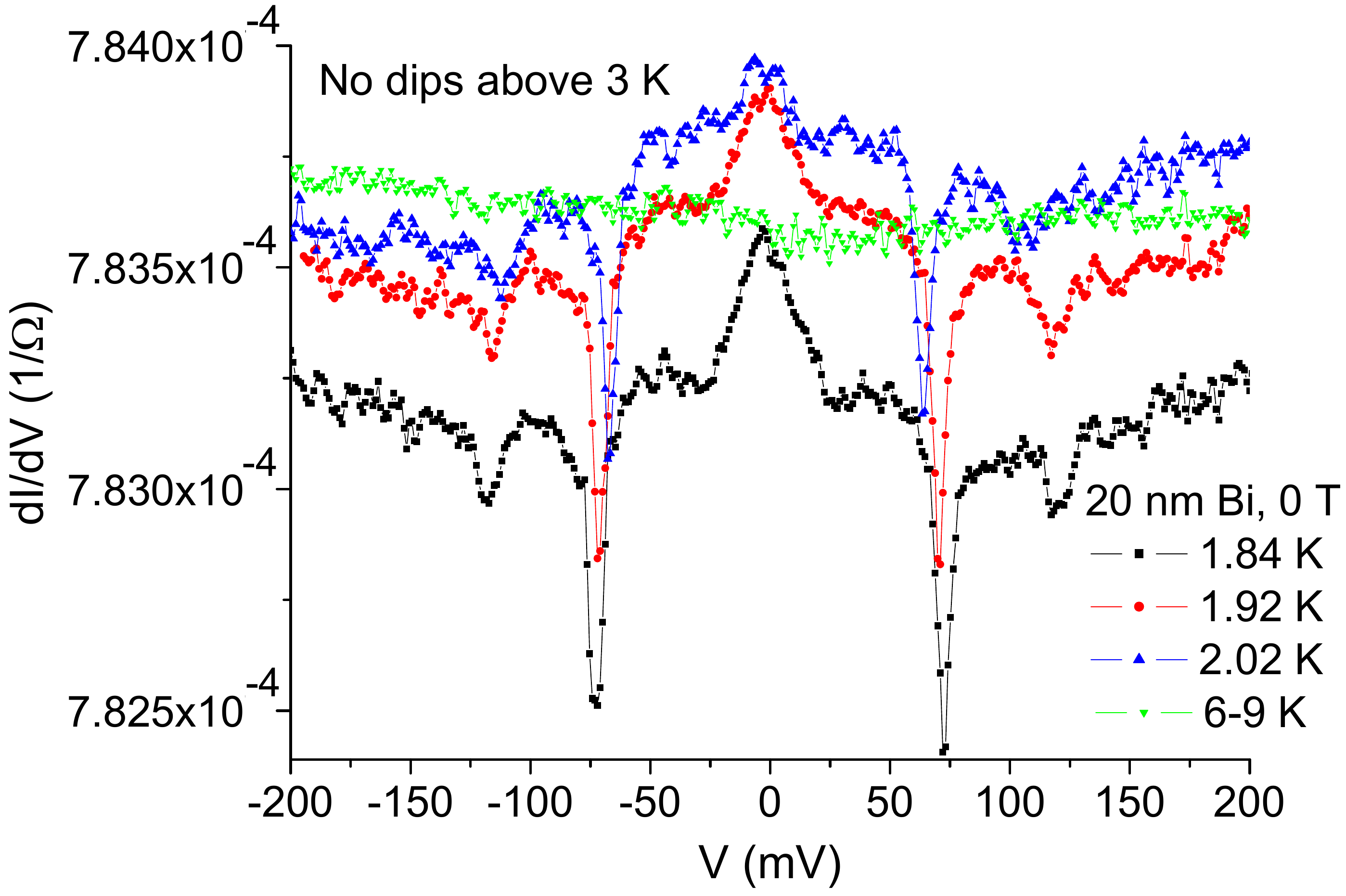}
\vspace{-0mm} \caption{\label{fig:epsart} (Color online) Conductance spectra at different temperatures of a point contact junction of gold in contact with a 20 nm thick $Bi$ film deposited on (100) $LaAlO_3$ wafer. No conductance dips were observed above about 3 K, and a typical background conductance is shown while heating between 6 and 9 K.}
\end{figure}

Since superconductivity in Bi is marginal, Bi is not superconducting in the bulk but only in small grains or on surfaces and interfaces \cite{Weitzel,Tian,Baring}, a reference 20 nm thick Bi film was prepared to test its superconducting properties. It was deposited at $\sim$50 K film surface temperature under vacuum on a (100) $LaAlO_3$ wafer, and its transport characteristics were measured. The 4-probe resistance versus temperature result in the range of 300-1.8 K showed quite a constant resistance with no clear signature of superconductivity. Superconductivity however was detected in 2-probe measurements of the conductance spectra of this film. Figs. 6 and 7 show typical results of such spectra and their behavior at different temperatures and under different magnetic fields. One can see that the conductance variations here are quite small, on the order of 0.1-0.2\% of the total conductance. Nevertheless, the conductance spectra of Figs. 6 and 7 show clear conductance dips which indicate that the critical current in the point contact junction was reached \cite{Sheet}. Similar dips have also been observed by Sasaki et al. in $Ag-Cu_xBi_2Se_3$ junctions (FIG. S4 of Ref. \cite{Ando}). Since these dips result from the critical current in the junctions, local superconductivity is clearly present in grains or islands in our Bi films, although no global superconductivity is observed. We note that the peak at low bias in Fig. 6 can also be due to reaching the critical current in weaker superconducting islands in the film, but can just as well result from Andreev scattering. The prominent dips of Fig. 6 disappear at temperatures above 3 K, and vanish under magnetic fields above 2 T as seen in Fig. 7. Thus, typical values of the superconducting critical temperature $T_c$ and the  critical field $H_{c2}$ of our Bi films are established, and will be useful in the present study of the proximity induced superconductivity in the $Bi-Bi_2Te_2Se$ bilayer and the $Bi_2Se_3$ films.\\

\begin{figure} \hspace{-20mm}
\includegraphics[height=9cm,width=13cm]{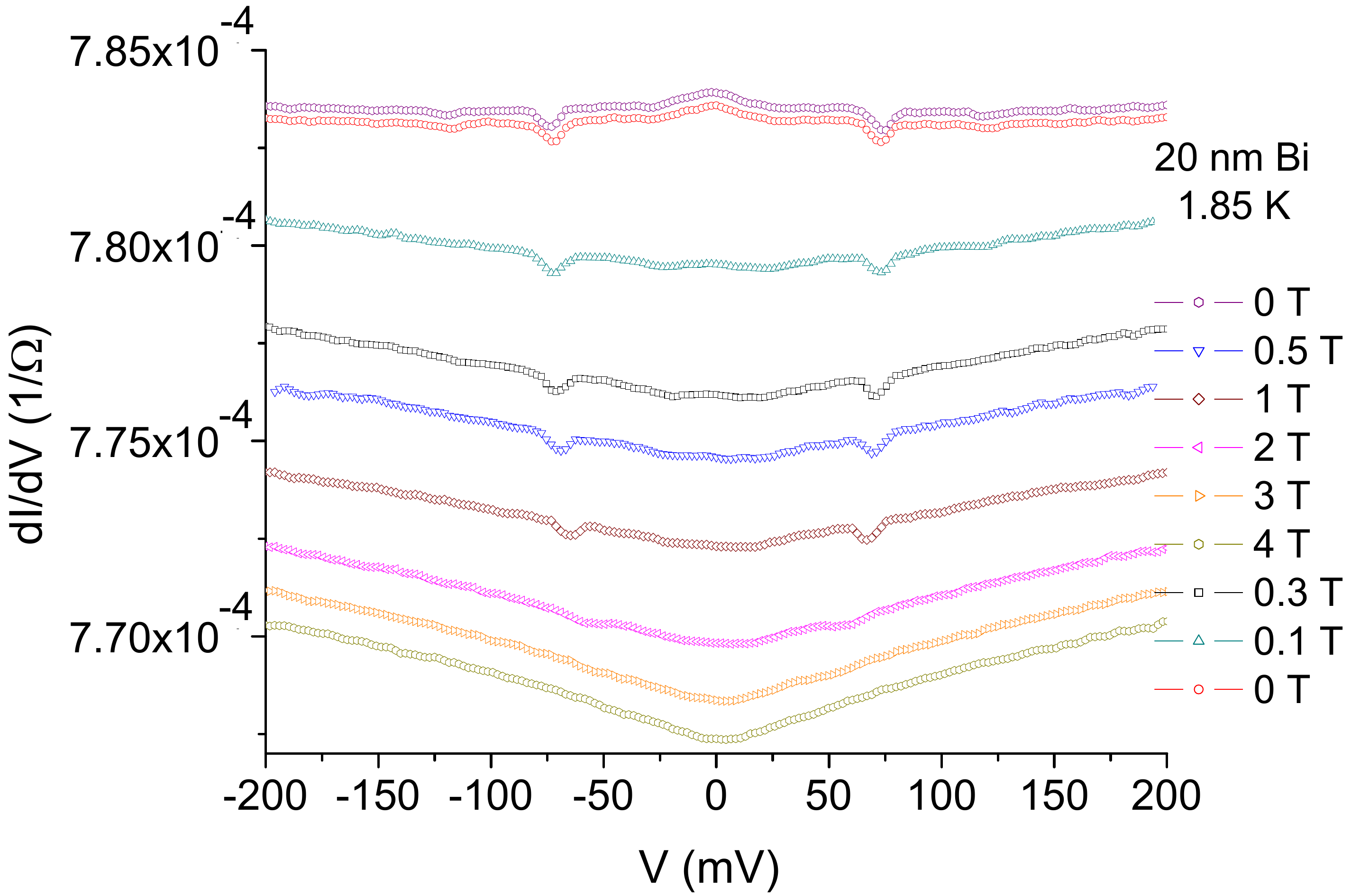}
\vspace{-0mm} \caption{\label{fig:epsart} (Color online) Conductance spectra at 1.85 K under different magnetic fields of the junction of Fig. 6.}
\end{figure}

 \begin{figure} \hspace{-20mm}
\includegraphics[height=9cm,width=13cm]{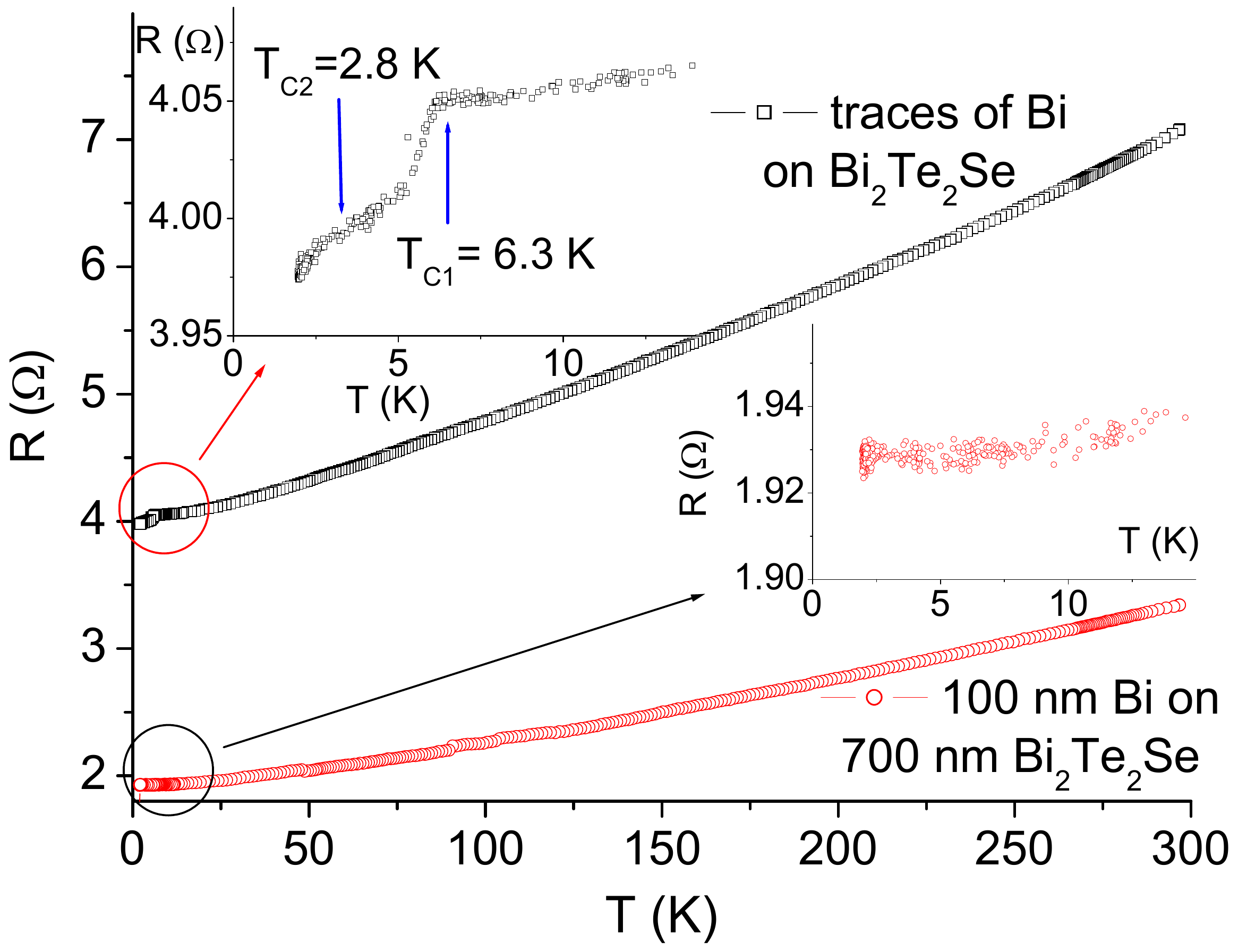}
\vspace{-0mm} \caption{\label{fig:epsart} (Color online) Resistance versus
temperature of two films. One with traces of Bi on top of 700 nm thick layer of
$Bi_2Te_2Se$ and the other of a bilayer of 100nm Bi on 700 nm $Bi_2Te_2Se$, together with zoom in on low temperatures in the insets.   }
\end{figure}

\section{Results and discussion}

\subsection{Point contacts of gold on a $Bi-Bi_2Te_2Se$ bilayer}

First we prepared a 700 nm thick $Bi_2Te_2Se$ (BTS) film on a whole wafer of (100) LAO. Then we shadow masked 1/3 of this wafer and continued with an \textit{in-situ} deposition of additional 100 nm thick layer of Bi. Fig. 8 shows representative 4-probe resistance versus temperature results of the two different parts of this wafer. Both curves show that these films are metallic. In the Bi/BTS bilayer with the thick Bi layer, no superconductivity was observed within the noise level of the measurement down to 1.8 K (see the bottom inset to Fig. 1). This result is consistent with the fact that bulk Bi is semi-metallic and not superconducting \cite{Weitzel,Tian,Baring}. Thus the 100 nm thick Bi layer in the Bi/BTS bilayer of Fig. 1 can be considered as bulk material. To further check this point, we deposited in a separate experiment a stand alone 100 nm thick film of Bi on (100) LAO and found that it showed no superconductivity down to 1.8 K either. However, a thinner, 10 nm thick Bi film on (100) LAO did show signatures of local superconductivity which appeared as a plateau in the resistance at 4-8 K on an insulating background resistance in one case, and as a sharp resistance drop at 2.3 K in another location on the wafer. The top inset of Fig. 8 shows that in the nominal BTS part of this film superconductivity does occur between superconducting islands with transition temperatures of 6.3 and 2.8 K. These islands are apparently due to either trace amounts of Bi that reached the surface under the shadow mask used when depositing the nearby thick Bi film, or to Bi segregation to the surface from the Bi rich BTS film itself due to the Se loss process. We confirmed this finding by energy dispersive spectroscopy (EDS) that showed the existence of a Bi rich layer on the surface of the BTS film, with a significant decrease of excess Bi in deeper layers of this film (see Fig. 4). We stress that although bulk Bi is not superconducting, on its surface, and in small clusters, nano-particles or islands it is superconducting \cite{Weitzel,Tian,Baring}. The two transition temperatures observed in the top inset of Fig. 8 can thus be attributed to the Bi islands (6.3 K) and to the proximity induced regions in the nearby TOI (2.8 K), respectively.\\

\begin{figure} \hspace{-20mm}
\includegraphics[height=9cm,width=13cm]{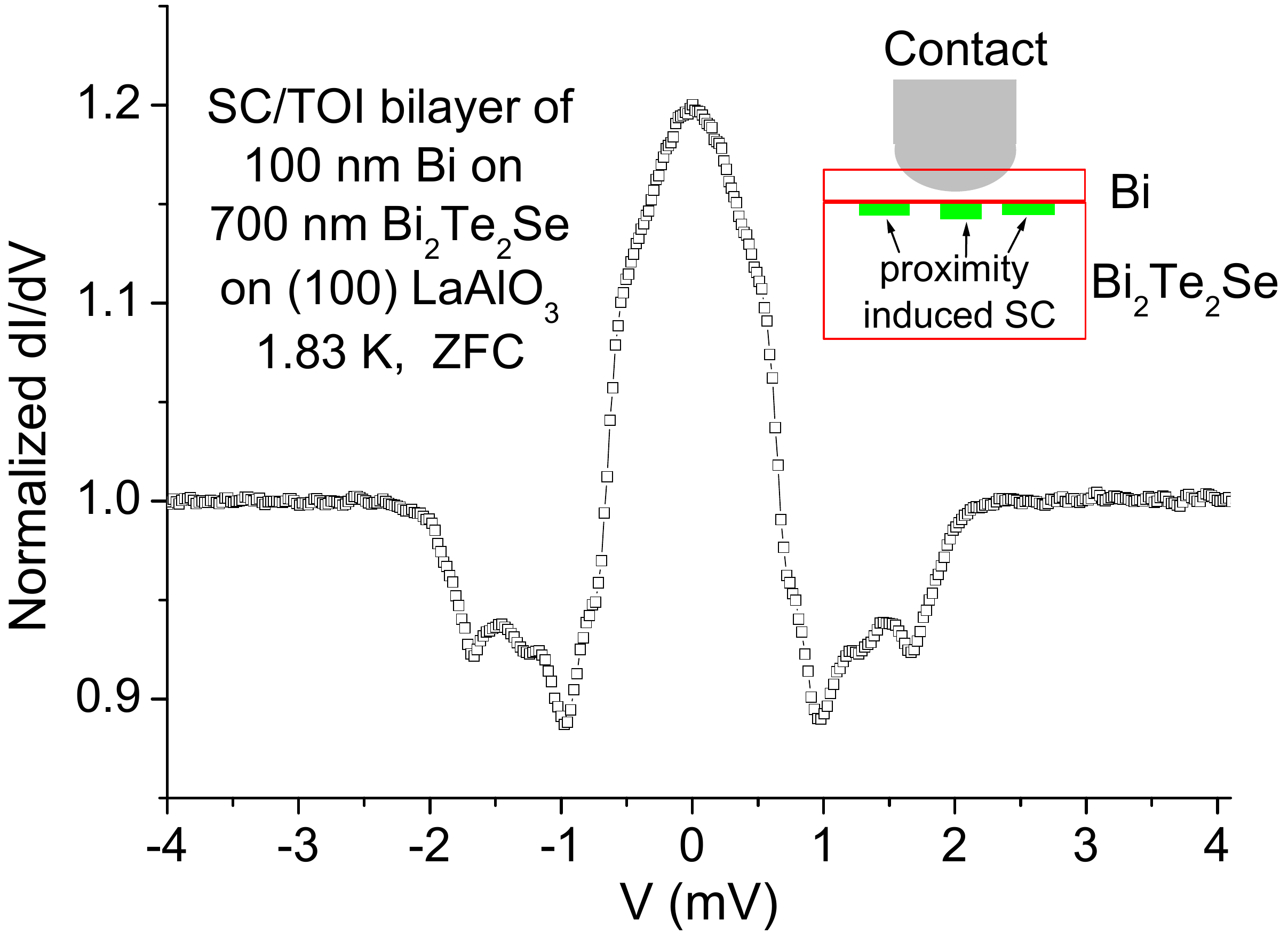}
\vspace{-0mm} \caption{\label{fig:epsart} (Color online) Normalized point contact junction conductance spectrum on the bilayer of Fig. 8. The inset shows a schematic drawing of the contact
configuration with the proximity induced superconducting islands close to the interface
of the Bi and $Bi_2Te_2Se$ layers. }
\end{figure}

A typical normalized conductance spectrum of one of the point contacts on the Bi/BTS bilayer is shown in Fig. 9 together with a schematic drawing of the contact geometry in the inset. This spectrum shows two prominent features: a pronounced ZBCP of about 2 mV width  and a tunneling-like gap at about $\pm$2 mV. The ZBCP is generally attributed to Andreev bound states due to unconventional superconductivity like in the cuprates where the d-wave order parameter changes sign, or to other type or orders such as $p_x+ip_y$ as proposed for the topological superconductors \cite{Tanaka,p-wave,Hao}. The spectrum in Fig. 2 is quite similar to the one calculated in Ref. \cite{Ando} for the $\Delta_4$ pair potential (odd parity with point nodes), but without the asymmetry. The additional dip features in the spectrum signify that the critical current in the superconducting Bi leads to the junction (contact) was reached which is typical in point contact measurements \cite{Sheet}. The fact that a few dips are observed, can be attributed to different values of the critical current in different superconducting Bi islands (grains) at the Bi/BTS interface. Comparing the overall conductance change in Fig. 9 ($\sim$30\%) and Fig. 6 (0.1-0.2\%), one can safely rule out the possibility that the robust ZBCP observed in Fig. 9 originates in the Bi film alone. We therefore conclude that the observed spectrum in Fig. 9 originates in the proximity induced superconducting zones near the interface of the bilayer as depicted in the inset to this figure.\\

\begin{figure} \hspace{-20mm}
\includegraphics[height=9cm,width=13cm]{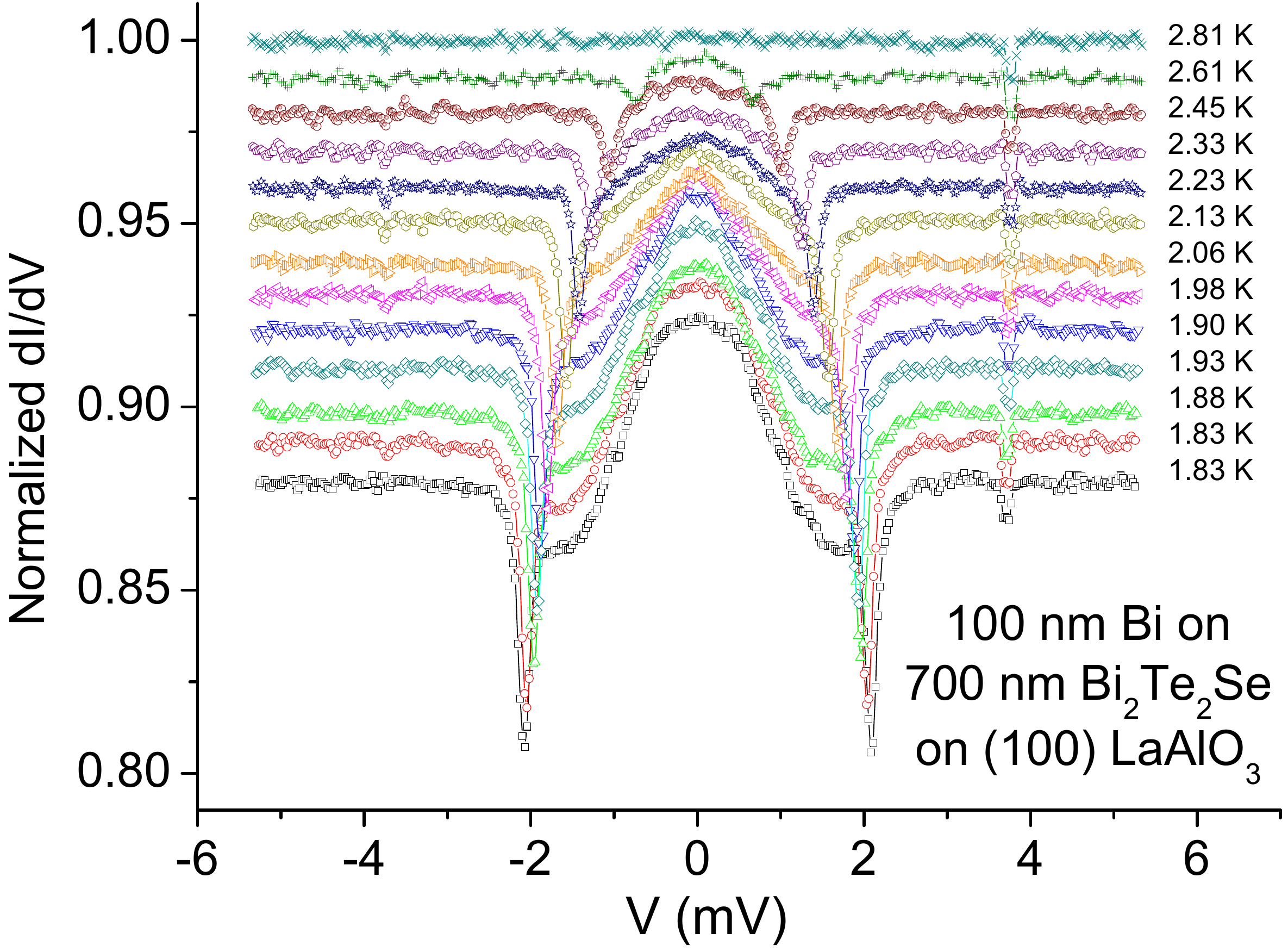}
\vspace{-0mm} \caption{\label{fig:epsart} (Color online) Normalized conductance spectra at different temperatures of another point contact junction on the $Bi-Bi_2Te_2Se$ bilayer of Fig. 8. For clarity, the spectra are shifted down compared to one another by 0.01 conductance units. }
\end{figure}

\begin{figure} \hspace{-20mm}
\includegraphics[height=9cm,width=13cm]{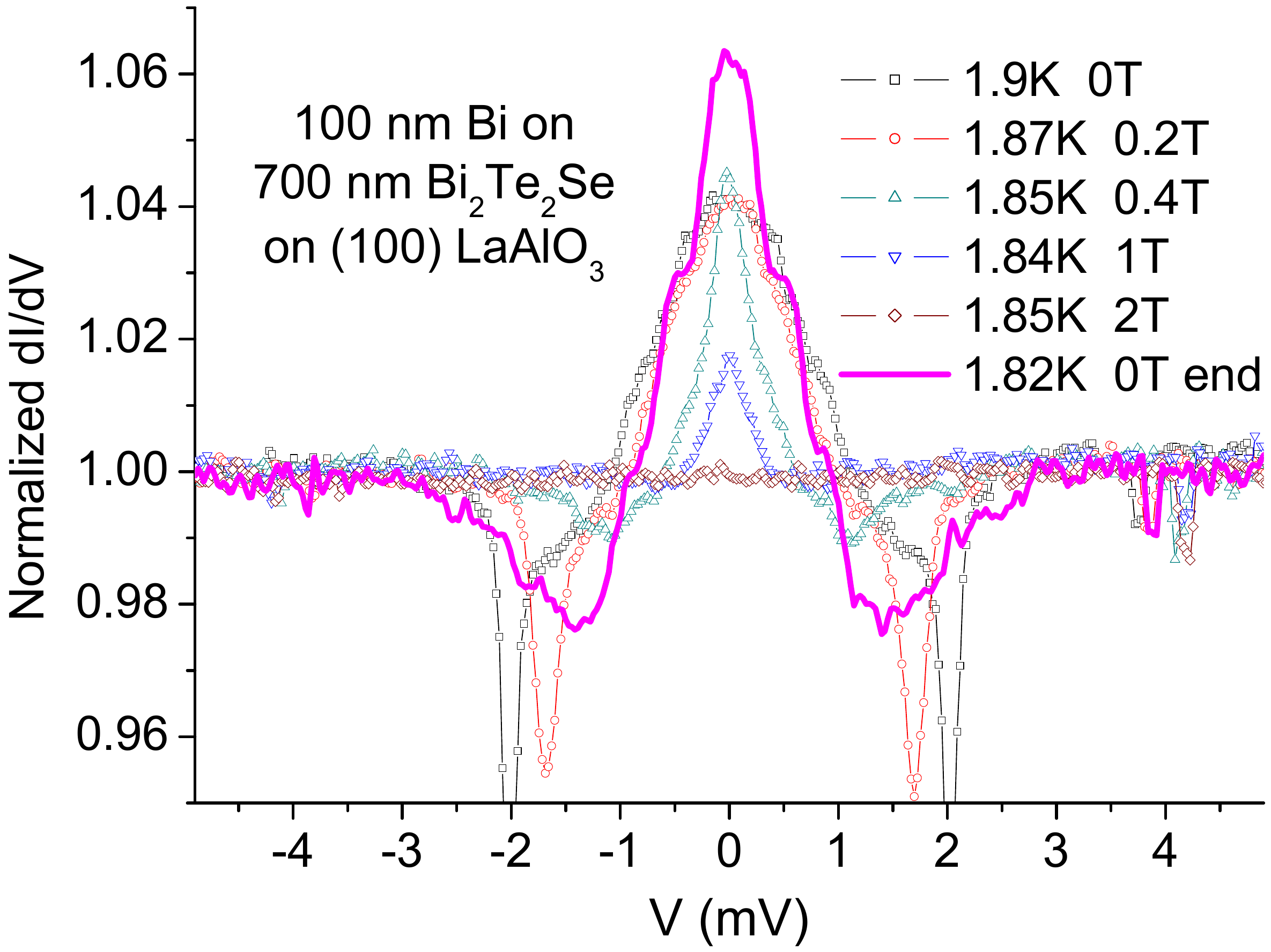}
\vspace{-0mm} \caption{\label{fig:epsart} (Color online) Normalized conductance spectra under different magnetic fields normal to the wafer of the same point contact junction of Fig. 10. }
\end{figure}

Fig. 10 shows the temperature dependence of the normalized conductance spectra of a different point contact on the same Bi/BTS bilayer. Here one observes at 1.83 K two very prominent critical current dips at about $\pm$2 mV which almost coincide with the tunneling-like gap feature and indicate that at least one dominant superconducting Bi island is present at this contact. With increasing temperature the ZBCPs decrease in height and narrow down, while the gap-like features close, and both disappear at 2.8 K. This temperature seems to coincide with the transition temperature of the proximity induced superconductivity of the junction in the TOI as seen in the top inset of Fig. 8. Fig. 11 shows the magnetic field dependence of the normalized conductance spectra of the same junction of Fig. 10 at low temperature (1.8-1.9 K). Basically, increasing the magnetic field has a similar effect to that of increasing temperature, but with a few important differences. First, the prominent critical current dips disappear above 0.2 T and do not reappear after field cycling to 2 T and back to zero field. This could be attributed to trapped flux in the superconducting Bi islands at the interface. Furthermore, these dips did not reappear after an additional temperature cycling to about 10 K and back to 1.8 K, while temperature  cycling without the application of fields did not affect them (the 1.9 K spectrum in Fig. 11 was obtained after the temperature cycling of Fig. 10). The spectrum under 0.4 T field is interesting in particular. It is composed of two contributions: a broad, kind of triangular, Andreev conductance peak of $\sim$0.8-1 mV width, and a much narrower ZBCP of $\sim$0.3-0.4 mV width on top of it. In this case, we identify the first peak as due to standard Andreev reflections below the gap energy, and the second peak as due to zero energy Andreev bound states. Generally, when the superconducting gap energy is small, it is hard to distinguish between these two contributions to the Andreev conductance, as is obvious from many other spectra in the present study. Besides the narrowing down and decay of the ZBCP and closing of the gap-like feature until their disappearance under a 2 T field, one can observe a clear enhancement of the ZBCP at 0 T after field cycling. This is a result of decreasing the density of states at high bias (1-2 mV, in the tunneling part of the spectrum) while conserving the integrated density of states. Furthermore, the ZBCP now is clearly composed of three contributions as can be seen by the steps in the spectrum at about  $\pm$0.3, $\pm$0.6 and $\pm$0.9 mV. These could be related to to different critical currents in different superconducting Bi islands as depicted in the inset to Fig. 9, or to the trapped vortex or vortices in these grains, but we have no way of supporting any of these hypothesis at the present time. The central part of the ZBCP is quite narrow now, only 0.5-0.6 mV, and the whole width of the ZBCP feature is of about 2 mV. These widths are in the range of typical surface Andreev bound states (SABS) such as found for helical edge states of a 2D TOI  \cite{Hsieh,Ando}. For odd parity pairing, some SABS can be considered as Majorana fermions, and therefore the presently observed ZBCP could be due to these particles. In addition, the observed ZBCP enhancement and its staircase shape with the three steps, is possibly related to trapped vortices. Theory predicts that at the interface between an s-wave superconductor (as the Bi here) and a strong TOI in the presence of a vortex, the resulting state supports Majorana bound states in the vortex core \cite{p-wave}. The spectrum in Fig. 11 at 0 T after field cycling could thus be due to Majorana fermions SABS resulting from the presence of a few (three?) trapped vortices. Clearly, further theoretical treatment of the data is needed to support any of the presently suggested interpretations of the results.\\

\begin{figure} \hspace{-20mm}
\includegraphics[height=9cm,width=13cm]{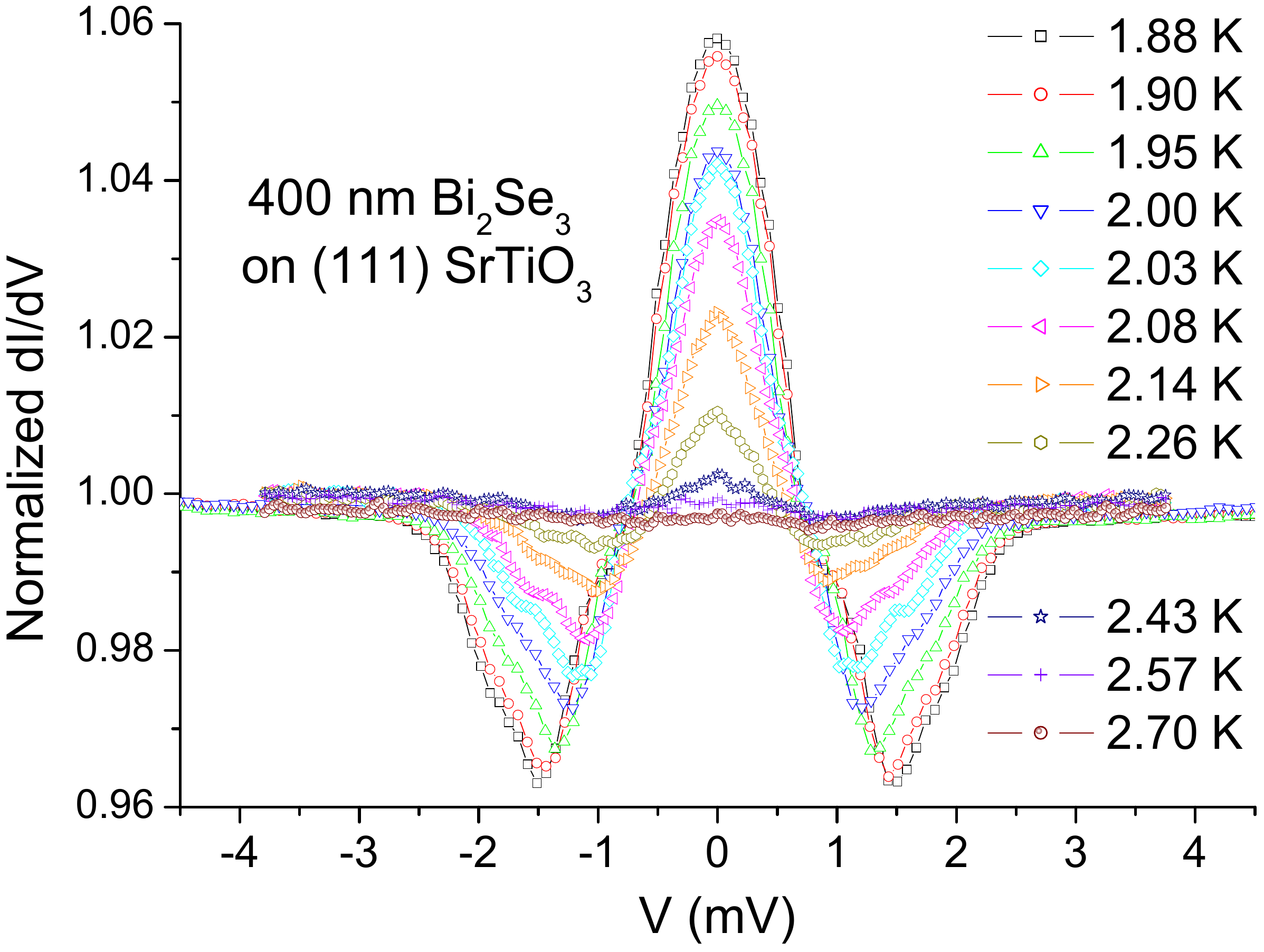}
\vspace{-0mm} \caption{\label{fig:epsart} (Color online) Normalized conductance spectra at different temperatures of a point contact junction on a $Bi_2Se_3$ film on (111) $SrTiO_3$ wafer. }
\end{figure}

\subsection{Point contacts of gold on $Bi_2Se_3$ films}

\begin{figure} \hspace{-20mm}
\includegraphics[height=9cm,width=13cm]{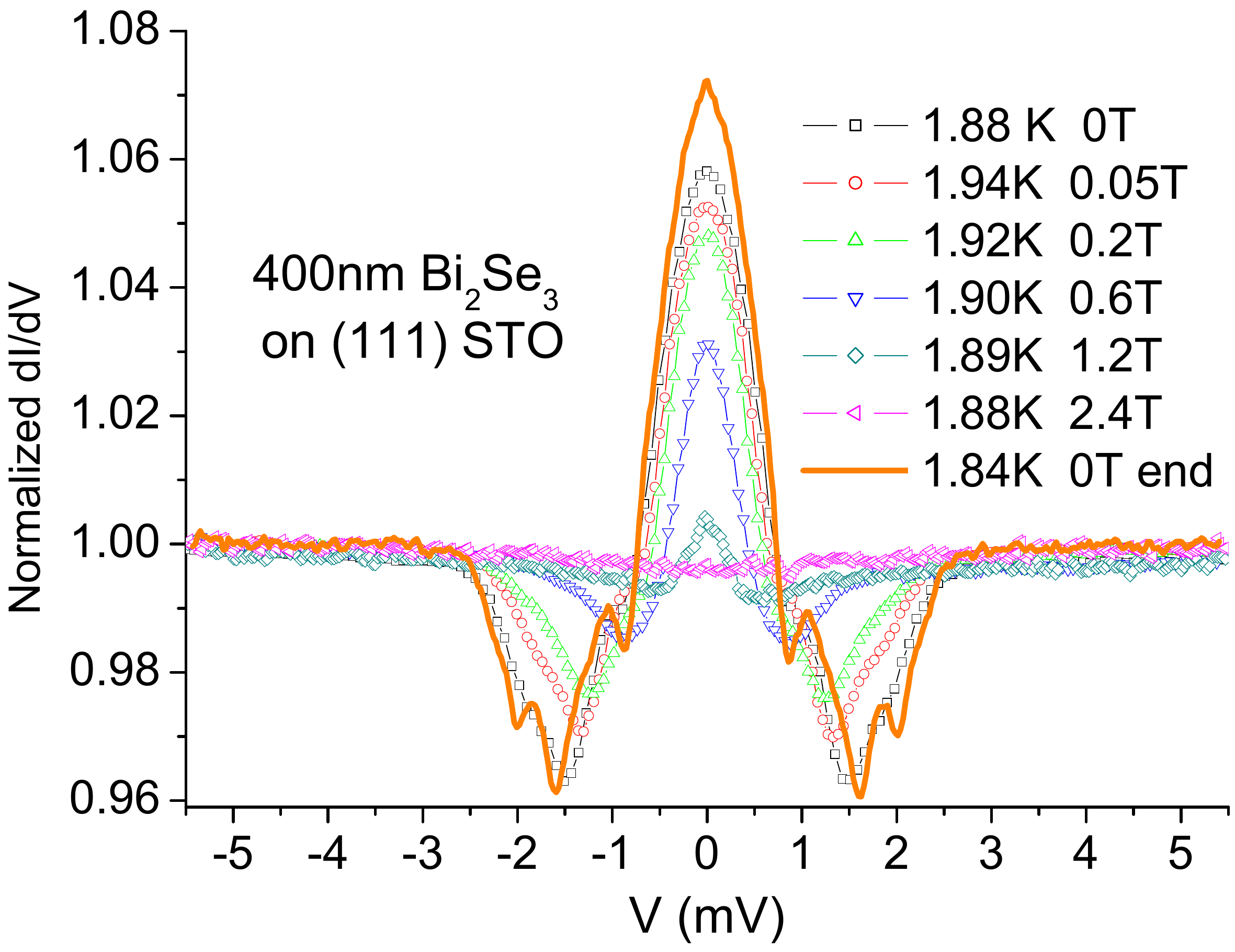}
\vspace{-0mm} \caption{\label{fig:epsart} (Color online) Normalized conductance spectra of the same point contact junction of Fig. 12 at $\sim$1.9 K under different magnetic fields normal to the wafer. }
\end{figure}

Figs. 12 and 13 show normalized conductance spectra of a point contact junction on the pure epitaxial $Bi_2Se_3$ film grown on (111) STO of Fig. 1. The data of Fig. 12 at different temperatures, and in Fig. 13 under different magnetic fields, is amazingly similar to that observed in Figs. 10 and 11, respectively, although no Bi layer was deposited on the chalcogenide film. The similarity of the results therefore suggests that either Bi segregation to the surface due to Se loss, or creation of Bi inclusions in the film is involved. EDS measurements on this film as seen in Fig. 4 substantiate the surface segregation scenario, but we can not rule out the Bi inclusions scenario at the present time. As before, the resistance versus temperature of this film showed no sign of superconductivity within the noise of the measurement. The symmetric dips in the spectra of Figs. 12 and 13 indicating the critical current effect as explained before \cite{Sheet}, necessitate the existence of superconducting grains in the junction. Thus the presence of local superconductivity here is established. Furthermore, the disappearance of the ZBCP at 2.7 K similar to the result of Figs. 10 and 6,  and under magnetic field of about 2 T similar to the results of Figs. 11 and 7, suggests that the same kind of Bi islands are involved. The enhancement of the ZBCP in Fig. 13 after the field cycling up to 2.4 T and back to 0 T is also similar to that observed in Fig. 11, but without the additional staircase structure of the ZBCP as in Fig. 11. Thus we conclude that if the Majorana fermion SABS interpretation of the ZBCP is correct for the Bi-BTS bilayer it should hold also for the blanket $Bi_2Se_3$ film. \\

\begin{figure} \hspace{-20mm}
\includegraphics[height=9cm,width=13cm]{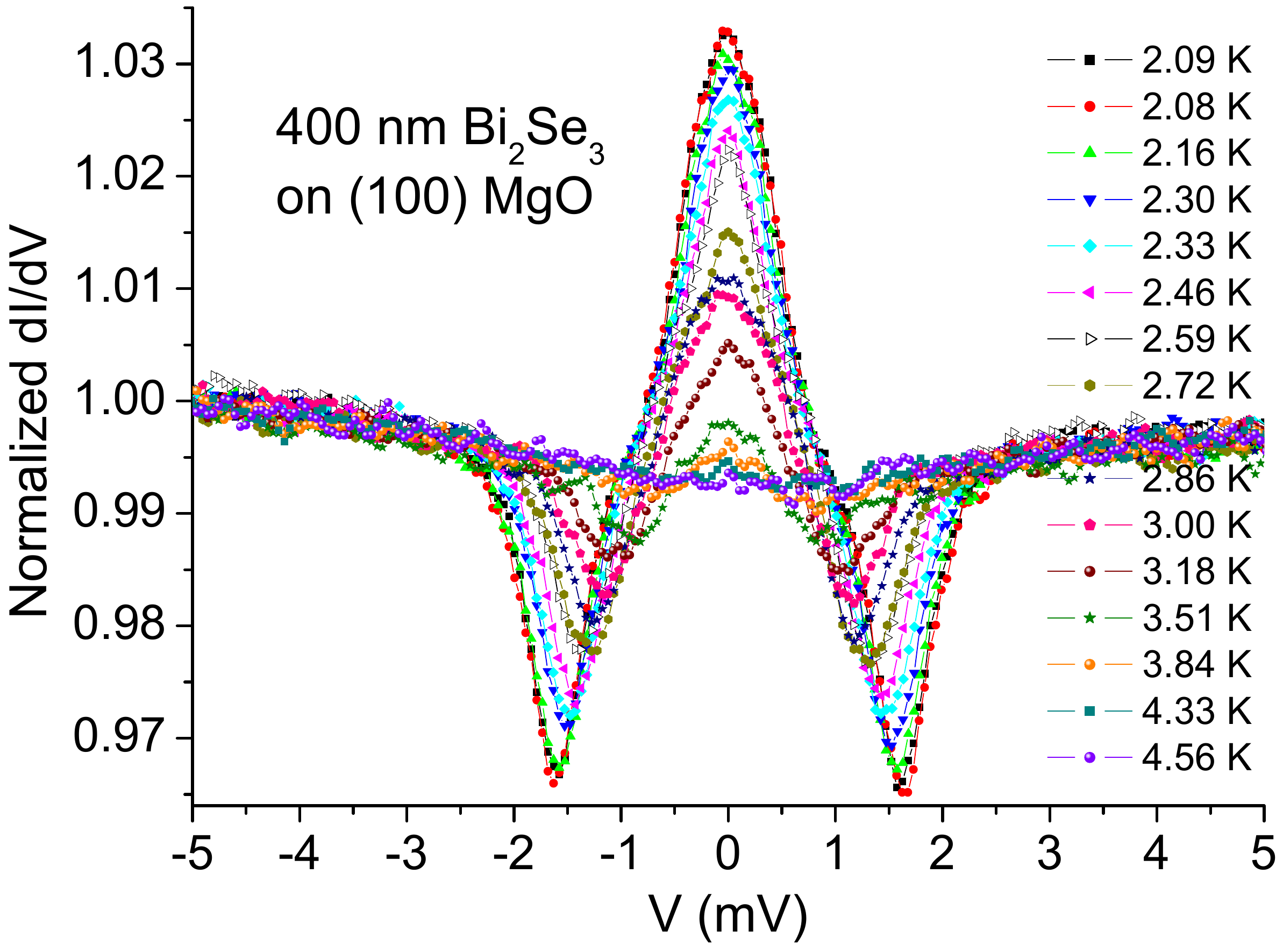}
\vspace{-0mm} \caption{\label{fig:epsart} (Color online) Conductance spectra of a point contact junction of gold and a $Bi_2Se_3$ film on (100) $MgO$ at zero field and different temperatures.}
\end{figure}

In order to check whether the effect of proximity induced superconductivity in $Bi_2Se_3$ films observed in Figs. 12 and 13  is robust and independent of the substrate, we repeated the conductance measurements on a similar film deposited on (100) $MgO$. The x-ray diffraction data of Fig. 2 shows that this film also has the same structure as that of the film on (111) $SrTiO_3$ (see Fig. 1). Figs. 14 and 15 show conductance spectra of a point contact junction of gold and this film. The data is shown at different temperatures in Fig. 14, and at $\sim$1.85 K under different magnetic fields in Fig. 15. The results are similar to those shown in Figs. 12 and 13 obtained on the $Bi_2Se_3$ film on (111) $SrTiO_3$. The features of the conductance spectra are basically the same for the pure films on both type of substrates, which gives further support for the reproducibility and robustness of the observed effect. It also proves that the effect has nothing to do with the substrate, or any possible local superconductivity which might originate in the $SrTiO_3$ wafer. Detailed inspection of the results of Fig. 14 shows that the ZBCP persists now up to about 4K, while it disappears already at about 3 K in Fig. 12. Fig. 15 shows that the critical current dips disappear at about 2 T, similar to the results in the bare Bi film of Fig. 7.  \\

\begin{figure} \hspace{-20mm}
\includegraphics[height=9cm,width=13cm]{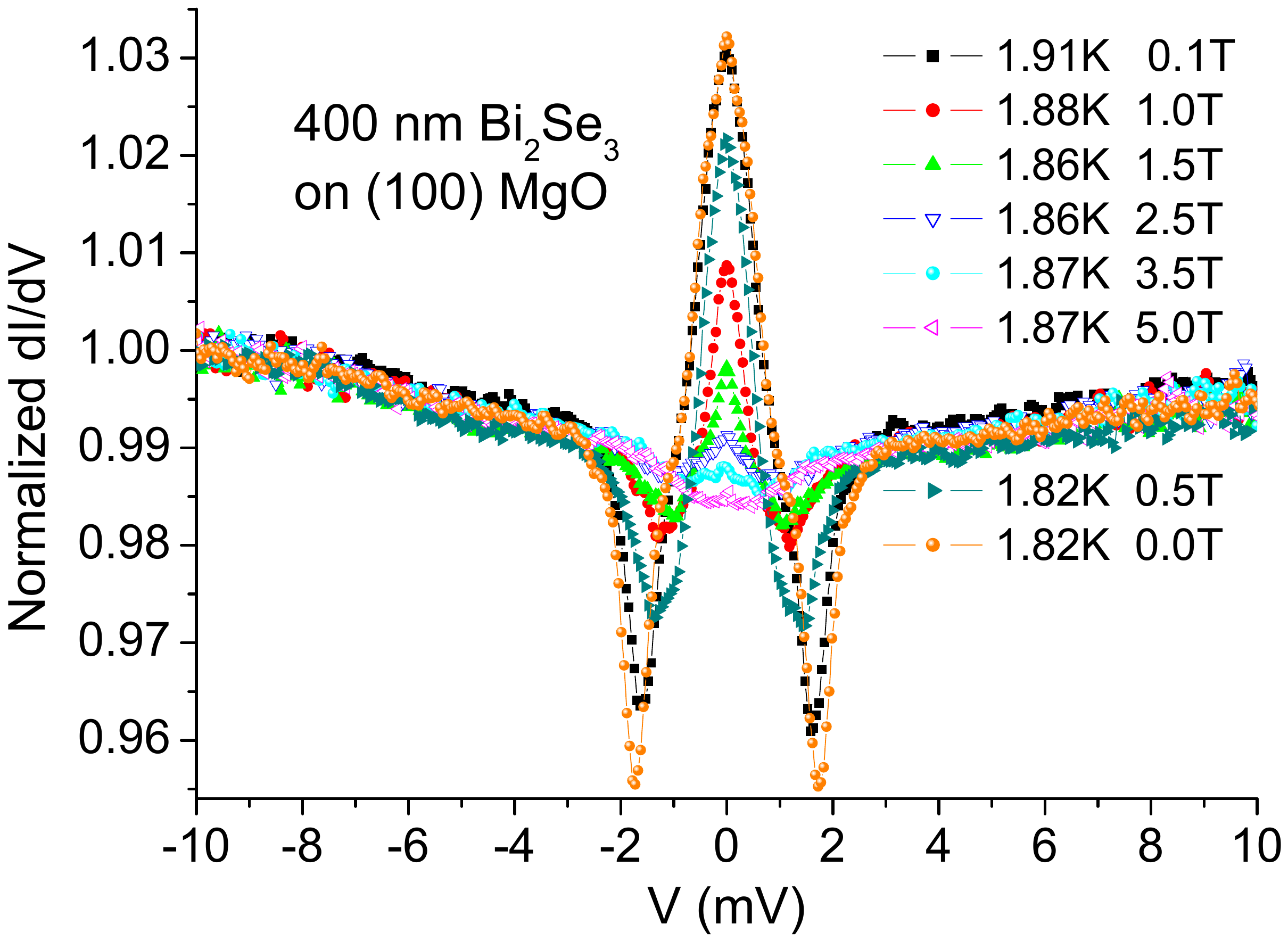}
\vspace{-0mm} \caption{\label{fig:epsart} (Color online) Conductance spectra of the same point contact junction of Fig. 14 at  $\sim$1.85 K under different magnetic fields.}
\end{figure}

And finally, some food for thought. Suppose one measures a ZBCP and dips in the conductance spectra of junctions of a normal metal or a superconductor with a topological insulator or a topological superconductor of the Bi-Se family. Then, we believe that it is essential to rule out the possibility that the observes effect originates in surface inhomogeneities or phase separation in the topological material, and result from a Bi induced proximity effect as seen here in the blanket $Bi_2Se_3$ films. These inhomogeneities if present,  could be due to exposure of the topological insulator to the ambient atmosphere, heating in a patterning process, intercalation chemistry, and so on.\\

In conclusions, point contact spectroscopy was performed on proximity induced superconducting regions in two topological insulators. Prominent zero bias conductance peaks were observed which were attributed to surface Andreev bound states, possibly related to Majorana fermions. The observed conductance dips were used as indicators of local superconductivity at the point contact regions even in the absence of global superconductivity in the films.\\

{\em Acknowledgments:}  This research was supported in part by the Israel Science Foundation, the joint German-Israeli DIP project and the Karl Stoll Chair in advanced materials at the Technion.\\

\bibliography{AndDepBib.bib}

\bibliography{apssamp}

\end{document}